\documentclass[twocolumn,showpacs,amsfonts,aps,prl,floatfix]{revtex4}
\usepackage{amsmath}
\usepackage{bm}
\usepackage{graphicx}
\usepackage{mathrsfs}
\usepackage{footmisc}
\usepackage{multirow}
\usepackage{graphicx}
\usepackage{booktabs, ctable}

\begin{document}

\title{ Detailed Rapidity Dependence of $J/\psi$ Production at energies available at the Large Hadron Collider}
\author{Baoyi Chen}
\affiliation{Department of Physics, Tianjin University, Tianjin 300350, China}
\date{\today}

\begin{abstract}
$J/\psi$ suppression with rapidity is studied with both (3+1) dimensional transport
equation and hydrodynamics in details at $\sqrt{s_{NN}}=2.76$ TeV Pb+Pb collisions.
In central rapidity region of the hot medium, $J/\psi$ nuclear modification factor shows weak rapidity dependence due to
the boost invariance of the medium. But in forward rapidity region for entire transverse momentum bin,
$J/\psi$ regeneration is strongly suppressed due to the decreasing charm pair cross section. This makes $J/\psi$ $R_{AA}(y)$ 
decrease with rapidity. 
The ratio $r_{AA}$ of mean transverse momentum square in AA collisions
to that in pp collisions increases strongly in forward rapidity region and 
is more sensitive to the $J/\psi$ production mechanisms.

\end{abstract}
\maketitle

\section{I. introduction}

$J/\psi$ is a tightly bound state consisting of charm and anti-charm quarks. 
Its dissociation temperature is much higher than the critical temperature $T_c$ of 
the deconfined phase transition. Therefore, $J/\psi$ suppression
has long been considered as a signature of the new state matter, the so-called 
Quark-Gluon-Plasma(QGP)~\cite{matsui}. Since then the production and suppression mechanisms of charmonium in QGP 
are widely studied~\cite{Thews:2000rj, Grandchamp:2002wp, Andronic:2003zv, Andronic:2011yq}.
The suppression from color screening of parton is supposed to be
stronger with the temperature increasing.
However, from the Au+Au data at the top colliding energy $\sqrt{s_{NN}}$= 200 GeV of
Relativistic Heavy Ion Collider (RHIC)~\cite{RHIC1}, it seems that $J/\psi$ suffer less suppression
in central rapidity region than in the forward
rapidity region of the hot medium, where the
former one corresponds to a much higher temperature than the latter one~\cite{Liu:2009wza,Zhao:2008pp,Krouppa:2015yoa}. 
This feature is much stronger at Large Hadron Collider (LHC). 
Even shadowing effect shows a strong rapidity dependence in pA collisions~\cite{Manceau:2013}, it's
almost rapidity independent in AA collisions~\cite{midpt1}. 
And $J/\psi$ regeneration~\cite{Andronic:2007bi} seems to be responsible for the rapidity 
dependence of $J/\psi$ nuclear modification factor $R_{AA}(y)$. In this work, 
I combine both (3+1) dimensional Boltzmann type transport equation for
charmonium evolution and (3+1) dimensional ideal hydrodynamics together, to 
describe detailed rapidity dependence of charmonium suppression, especially 
in the forward rapidity region at $\sqrt{s_{NN}}=2.76$ TeV Pb+Pb collisions.

In Section II, I introduce the transport model for charmonium suppression and regeneration 
in QGP. In Section III, I give detailed discussions about charmonium initial distribution and cold nuclear 
matter effects including shadowing effect and Cronin effect. 
I compare the theoretical calculations with the experimental data, 
and discuss the different rapidity dependence of parameters 
in the transport model in Section IV, and summarize the work in Section V.  

\section{II. transport model}
As charmonium is so heavy to be thermalized in the deconfined matter,
I use a Boltzmann-type transport equation to describe the evolution of their phase space
distribution $f_{\Psi}({\bf p},{\bf x},t|{\bf b})$~\cite{trans1, Liu:2009gx},
{\footnotesize
\begin{equation}
\left[\cosh(y-\eta){\frac{\partial}{\partial\tau}}+{\frac{\sinh(y-\eta)}{\tau}}{\frac{\partial}{\partial
\eta}}+{\bf v}_T\cdot\nabla_T\right]f_\Psi
=-\alpha_\Psi f_\Psi+\beta_\Psi
\label{tran}
\end{equation}
}where ${\Psi}$ = ($J/\psi$, $\chi_c$, $\psi^\prime$). The contributions from the feed-down of excited states
$\chi_c$ and $\psi^\prime$ to the final prompt $J/\psi$ are 30\% and 10\% respectively. 
$\tau=\sqrt{t^2-z^2}$ is local proper time, and 
$y={1\over 2}\ln[(E+p_z)/(E-p_z)]$ and $\eta={1\over 2}\ln[(t+z)/ (t-z)]$ are rapidities in momentum and coordinate spaces. 
${\bf \nabla}_T$ indicates two
dimensional derivative in transverse coordinate space. ${\bf b}$ is the
impact parameter of two colliding nuclei. Due to interactions with the hot medium,
charmonium distribution $f_{\Psi}({\bf p},{\bf x},t|{\bf b})$ changes with time which is 
described in the first term at L.H.S of Eq.(\ref{tran}). 
The second term represents leakage effect, where charmonium with large
transverse velocity ${\bf v}_T={\bf p}_T/E_T$ tend to suffer less suppression
by free streaming from the hot medium, $E_T=m_T=\sqrt{m_\Psi^2+p_T^2}$ is charmonium transverse energy. 
When charmonium moving in the QGP, 
it may be dissociated by 
inelastic collisions with gluons, and the dissociation rate is described by the term $\alpha_\Psi$ in Eq.(\ref{tran}), 
{\footnotesize
\begin{equation}
\label{Cdfactor}
\alpha_\Psi ={1\over 2E_T} \int {d^3{\bf k}\over {(2\pi)^32E_g}}\sigma_{g\Psi}({\bf p},{\bf k},T)4F_{g\Psi}({\bf p},{\bf k})f_g({\bf k},T)
\end{equation}
}where $E_g$ is gluon energy, 
$F_{g\Psi}$ is the flux factor. $f_g= d_g/(e^{p_g\cdot u\over T}-1)$ is the gluon thermal distribution with the 
degeneracy $d_g=16$. 
$u_\mu$ is four velocity of the fluid. 
Due to strong color screening of QGP, 
the binding energy of $J/\psi$ is reduced. 
The difference between quasifree dissociation ($g,q+c\rightarrow g,q+c$) 
and photo-dissociation ($g+J/\psi\rightarrow c+\bar c$) is large at high 
temperature~\cite{Grandchamp:2001pf, Grandchamp:2003uw, Zhao:2010nk}. However, their values 
are close to each other in the temperature region: $0.2\sim 0.3$ GeV~\cite{Grandchamp:2001pf}. 
Therefore, in this temperature region, 
one can employ the 
inelastic scattering cross section between $J/\psi$ and gluon 
$\sigma_{g\Psi}$(0) in vacuum~\cite{OPE1,OPE2}, and obtain its value at finite temperature with geometry scale 
$\sigma_{g\Psi}(T)=\langle r_\Psi^2\rangle(T)/\langle r_\Psi^2\rangle(0) \times \sigma_{g\Psi}(0)$. 
The mean 
radius square $\langle r_\Psi^2\rangle$ can be obtained 
by solving Schr\"odinger equation, with charmonium potential to be its internal energy V=U~\cite{radius2}. 
The divergence of mean radius of charmonium at $T\rightarrow T_d$ 
indicates the melting of bound states. The value of $T_d(J/\psi)$ is slightly above the 
region where photo-dissociation of $J/\psi$ can be employed. At $T>T_d$, All charmonium is melt in this approach. 
Elastic collisions between charmonium and the hot medium have been checked~\cite{elastic1}
and found to be negligible for $J/\psi$ $p_T$-integrated observables.

During the evolution of QGP, uncorrelated $c$ and $\bar c$ from different $c\bar c$ pairs may 
combine to form a new charmonium, which is called ``regeneration''~\cite{Thews:2000fd, Grandchamp:2001pf}. 
The regeneration rate is represented by the term $\beta_\Psi$ in Eq.(\ref{tran}), 
and it can be obtained from $\alpha_\Psi$ by detailed balance~\cite{rege1}.
The yield of regenerated charmonium is 
roughly proportional to $N_{c\bar c}^2$. 
Therefore at high colliding energies as LHC where $N_{c\bar c}$ is large, 
the yield of regenerated charmonium 
becomes important (even dominates) in the final yields of charmonium~\cite{Zhao:2011cv,Du:2015wha,trans2}. 
Both initially produced charmonium and 
regenerated charmonium suffer suppression in QGP. 
At LHC colliding energies, QGP dominates the charmonium evolution and hadron phase contribution can be neglected. 

Charm quark takes some time to reach thermal momentum distribution, due to its large mass~\cite{Zhao:2012gc}. 
In the beginning of QGP evolution, charm quark momentum distrbution should be non-equilibrium. 
At this time, the initial temperature of QGP is much higher ($T_{\mathrm{QGP}}>T_d(J/\psi)$), 
where charmonium regeneration 
can not happens (all the charmonium is melt at $T>T_d$). Only when the temperature of QGP drops below $T_d(J/\psi)$, 
the regenerated charmonium can survive. At the time where charmonium regeneration 
happens, charm quark should be closer to the 
thermal momentum distribution, compared with their initial distribution. 
Also, D meson elliptic flow is comparable with that of light hadrons~\cite{ctherm1, Greco:2003vf}, 
one can take the thermal momentum distribution of charm quarks in QGP. 
The non-equilibrium of charm momentum distribution will suppress the yields of 
regenerated charmonium~\cite{Zhao:2010nk}. At 
forward rapidity bin, the regeneration will be more suppressed by this non-equilibrium effect. And the decreasing tendency 
of solid line in Fig.\ref{pt0} will be slightly enhanced. 

With the conservation of total number of charm pairs during the evolution of QGP, 
its spatial density evolution satisfies the conservation law,
${\partial _\mu}(\rho_cu^\mu)=0$. 
Charm quark initial spatial distribution is obtained by nuclear geometry,
{\footnotesize 
\begin{equation}
\label{dnsty}
\rho_c({\bf x}_T, \eta,\tau_0) = {{d\sigma^{pp}_{c{\bar c}}} \over d\eta}{ T_A({\bf x}_T)T_B({\bf x}_T-{\bf b})\cosh(\eta)\over \tau_0}
\end{equation}
}where $d\sigma^{pp}_{c{\bar c}}/d\eta$ is rapidity distribution of charm pairs produced in pp collisions,
$T_A$ and $T_B$ are thickness functions of two colliding nuclei. It is defined
as $T_{A(B)}({\bf x}_T)=\int_{-\infty}^{\infty} dz\rho_{A(B)}({\bf x}_T,z)$, and $\rho_{A(B)}({\bf x}_T,z)$ is Woods-Saxon nuclear
density.

\section{III. cold nuclear matter effects}
Initially produced charmonium is included as an input of Eq.(\ref{tran}). 
With cold nuclear matter effects, charmonium initial distribution at $\tau=\tau_0$ in Pb+Pb collisions is,
{\footnotesize
\begin{align}
f_\Psi^{\mathrm{init}}({\bf x},{\bf p}|{\bf b}) &= {(2\pi)^3\over E_T\tau_0}\int dz_A dz_B\rho_A({\bf x}_T, z_A)\rho_B({\bf x}_T-{\bf b}, z_B) \nonumber \\ 
&\times R_g^{Pb}(x_1, Q^2, {\bf x}_T) R_g^{Pb}(x_2, Q^2, {\bf x}_T-{\bf b})  \nonumber \\
&\times f_\Psi^{pp+\mathrm{cronin}}({\bf x}_T,{\bf p},z_A, z_B|{\bf b}) 
\label{fPsiInit}
\end{align}
}$R_g^{Pb}(x_{1,2},Q^2,{\bf x}_T)$ is the gluon shadowing function from EKS98 package with transverse 
coordinate dependence~\cite{Eskola:1998df,Vogt:2010aa}. 
The relations between charmonium momentum and ($x_{1,2},Q^2$) are $x_{1,2}=(m_T/\sqrt{s_{NN}})\exp(\pm y)$, 
and $Q^2=m_T^2$~\cite{Vogt:2004dh}. When two nuclei collide with each other, 
gluons may obtain extra energy by multi-scatterings with nucleons before fusing into a charmonium. 
This effect (called Cronin effect) results in the $p_T$-broadening of 
charmonium $\Delta \langle p_T^2\rangle =\langle p_T^2\rangle_{pA}-\langle p_T^2\rangle_{pp}$~\cite{cronin1}. It 
can be implemented by Gaussian smearing over the power-law $p_T$-spectra from pp 
collisions~\cite{Zhao:2008pp}, 
{\footnotesize
\begin{align}
f_\Psi^{pp+\mathrm{cronin}}={1\over \pi \Delta \langle p_T^2\rangle}\int d{\bf p}_T^{\prime 2}
\exp(-{{\bf p}_T^{\prime 2}\over \Delta \langle p_T^2\rangle}) f_{\Psi}^{pp}(|{\bf p}_T-{\bf p}_T^\prime|, p_z)
\end{align}
}$f_\Psi^{pp}$ is the charmonium momentum distribution in free proton-proton collisions. 
$\Delta \langle p_T^2\rangle=a_{gN} \cdot \langle l\rangle$ where $\langle l\rangle$ 
represents the mean nuclear path length before gluons 
fusing into $\Psi$~\cite{Hufner:2001tg}, and the coefficient is 
extracted as $a_{gN}=0.15\ \mathrm{GeV^2/fm}$~\cite{trans2,Thews:2005vj}. 
For the charmonium momentum distribution in free pp collisions, it can be parametrized as 
{\footnotesize
\begin{align}
&{d^2\sigma_\Psi^{pp}\over dyp_Tdp_T} = f_\Psi^{\mathrm{Norm}}(p_T|y)\cdot {d\sigma_\Psi^{pp}\over dy}(y) \\
&f_\Psi^{\mathrm{Norm}}(p_T|y)={(n-1)\over \pi (n-2)\langle p_T^2\rangle_\Psi^{pp}(y)}
(1+{p_T^2\over (n-2)\langle p_T^2\rangle_\Psi^{pp}(y)})^{-n}
\end{align}
}with $n=4.0$. In this work, I focus on the rapidity dependence of charmonium yield and momentum distribution. 
The rapidity dependence of charmonium yield is characterized 
by ${d\sigma_\Psi^{pp}\over dy}(y)$. $f_\Psi^{\mathrm{Norm}}(p_T|y)$ is the charmonium 
transverse momentum distribution at different rapidity, normalized to unit. It depends on rapidity through 
the mean transverse momentum square $\langle p_T^2\rangle_\Psi^{pp}(y)$.
At PHENIX, $\langle p_T^2\rangle_\Psi^{pp}(y)$ is close to the form~\cite{phenix1} 
{\footnotesize
\begin{equation}
\langle p_T^2\rangle_\Psi^{pp}(y)=\langle p_T^2\rangle_\Psi^{pp}|_{y=0}\times [1-({y\over y_{\mathrm{max}}})^2]
\label{ydepend}
\end{equation}
}where $y_{\mathrm{max}}=\mathrm{cosh^{-1}}(\sqrt{s_{NN}}/(2m_{T}))$ 
is the maximum rapidity of charmonium
in pp collisions. 
At the maximum rapidity, charmonium transverse momentum is small (close to zero) 
compared with $m_\Psi$, 
so one can take the approximation $m_T\approx m_\Psi$ in the calculation of $y_\mathrm{max}$. 
At LHC $\sqrt{s_{NN}}=2.76$ TeV pp collisions, I also take this parametrization and
$\langle p_T^2\rangle_\Psi^{pp}|_{y=0}=10\ \mathrm{(GeV/c)^2}$ for $J/\psi$~\cite{Abelev:2012kr}.
Initial distributions of excited states ($\chi_c$, $\psi^\prime$) are taken the same as $J/\psi$'s for simplicity.
As charm quarks tend to be thermalized in QGP, final charmonium observables are almost not affected by 
charm quark initial momentum distribution. From EKS98 package, shadowing effect roughly reduces 
the total yield of charm pairs by around $20\%$, and this effect also shows 
weak rapidity dependence in Pb+Pb collisions~\cite{Vogt:2010aa}. 

\begin{figure}[htb]
{\includegraphics[width=0.48\textwidth]{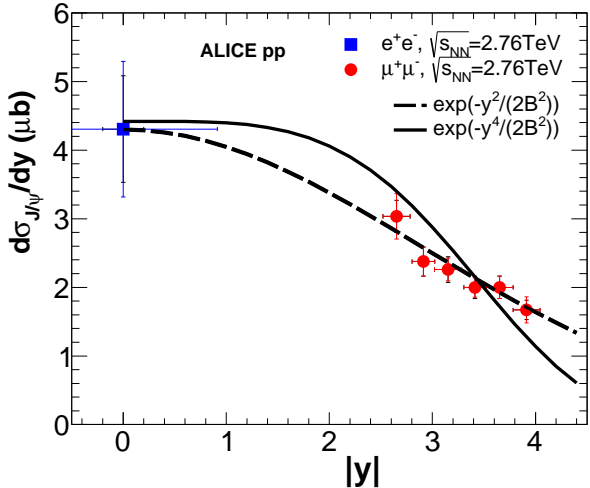}
\caption{(Color online) Differential inclusive $J/\psi$ production cross 
section at $\sqrt{s_{NN}}=2.76$ TeV pp collisions, fitted by exponential 
functions ${d\sigma_{J/\psi}^{pp}\over dy}=A\exp(-y^n/(2B^2))$. Solid and dashed lines 
correspond to $n=4$ and $n=2$ respectively. 
The parameters $A$ and $B$ are listed in table \ref{TabParamRap}. Data is from ALICE Collaboration~\cite{Abelev:2012kr}. 
}
\label{figJpsiSigma}}
\end{figure}

The initial inclusive $J/\psi$ production cross section with rapidity in $\sqrt{s_{NN}}=2.76$ TeV pp 
collision is measured by ALICE Collaboration at central
and forward rapidity regions~\cite{Abelev:2012kr}, it is parameterized in Fig.\ref{figJpsiSigma}. 
Considering total prompt charmonium yield is about 90\% of
the inclusive yield in pp collisions, the cross section of prompt charmonium can be taken as $0.9\times {d\sigma^{pp}_{J/\psi}\over dy}$.
In each $p_T$ bin, the fractions of non-prompt $J/\psi$ from B decay contribution can be extracted from~\cite{Chatrchyan:2012np}. 
\begin{table} [h]
\normalsize
\centering
\caption{parameters of ${d\sigma_{J/\psi}^{pp}\over dy}(y)$ in Fig.\ref{figJpsiSigma}}
\label{TabParamRap}
\begin{tabular}{|p{1.8cm}<{\centering}|*{4}{p{1.8cm}<{\centering}|}} 
\hline
           &  $A$ ($\mu b$) & $B$ \\
\hline
solid line & 4.42 & 9.71\\
\hline
dashed line & 4.30 & 2.88\\
\hline
\end{tabular}
\end{table}

At LHC colliding energies, the yield of initially produced charmonium are strongly suppressed 
by QGP due to color screening and gluon dissociation. The regenerated 
charmonium from charm and anti-charm quark recombination 
dominates the final yield of $J/\psi$ in the low $p_T$ bins. 
The yield of regenerated charmonium is roughly proportional to the 
square of charm quark density. Therefore different parameterization of ${d\sigma_{J/\psi}^{pp}\over dy}(y)$ and 
${d\sigma_{c\bar c}^{pp}\over dy}(y)$ will result in different shape of $R_{AA}^{J/\psi}(y)$. In order to 
consider the uncertainties in Fig.\ref{figJpsiSigma}, 
I give two scenarios of ${d\sigma_{J/\psi}^{pp}\over dy}(y)$ with slowly decreasing (dashed line) 
and rapidly decreasing (solid line) tendencies 
in the region $2<|y|<4$ (this is the region where $R_{AA}^{J/\psi}(y)$ shows strong decreasing tendency in Fig.\ref{pt0}). 
Even dashed line seems to fit data better in Fig.\ref{figJpsiSigma}, 
however, pQCD calculations indicate that at LHC colliding energies, 
the flat region of ${d\sigma_{c\bar c}^{pp}\over dy}$ can be as large 
as $|y|<3$ at $\sqrt{s_{NN}}=5.5$ TeV pp collisions~\cite{Andronic:2006ky}, 
which supports the solid line in Fig.\ref{figJpsiSigma}. 
Both situations will be considered below. 
I use the same parametrization for charm pair production cross section, 
${d\sigma_{c\bar c}^{pp}\over dy}(y)={S}\times {d\sigma_{J/\psi}^{pp}\over dy}(y)$, 
and $S$ is a constant. 
Only the data of total charm pair cross section $\sigma^{\mathrm{total}}_{c{\bar c}}= 4.8^{+4.66}_{-2.76}$ mb
and ${d\sigma^{pp}_{c{\bar c}}\over dy}(|y|<0.5)$ (with large error-bar) 
are available from ALICE Collaboration~\cite{cpair1}. Lack of more constraint, 
I fit the $J/\psi$ nuclear modification factor $R_{AA}^{J/\psi}(N_p)$ 
in central rapidity region~\cite{trans2} by setting ${d\sigma^{pp}_{c{\bar c}}\over dy}(y=0)=0.85$ mb, which is also close to 
the cross section from FONLL model~\cite{trans2,Cacciari:2001td}. 
It results in $S=0.85\mathrm{mb}/A$ (A given in table \ref{TabParamRap}). 

The inclusive nuclear modification factor of charmonium is defined as 
{\footnotesize
\begin{align}
\label{raaEq}
R_{AA}^{\Psi}(\mathrm{inclusive}) &={N_{AA}^{\Psi}(\mathrm{init})+ N_{AA}^{\Psi}(\mathrm{rege}) +N_{AA}^{\mathrm{B\rightarrow \Psi}}
\over (N_{pp}^{\Psi} +N_{pp}^{B\rightarrow \Psi})\cdot N_{coll}} \nonumber \\
&=R_{AA}^{\mathrm{init}} +R_{AA}^{\mathrm{rege}} +R_{AA}^{\mathrm{B\rightarrow \Psi}}
\end{align}
}$N_{AA}^{\Psi}(\mathrm{init},\mathrm{rege},\mathrm{B\rightarrow \Psi})$ is the yield of $\Psi$ 
from (initial production, regeneration, B hadron decay) respectively in nucleus-nucleus collisions. 
$N_{pp}^{\Psi}$ is the yield of $\Psi$ in pp collisions.
$N_{coll}$ is the number of binary collisions at fixed impact parameter ${\bf b}$. 
In the Eq.(\ref{raaEq}), all the denominators of three terms ($R_{AA}^{\mathrm{init}}$, 
$R_{AA}^{\mathrm{rege}}$, $R_{AA}^{\mathrm{B\rightarrow \Psi}}$) are $(N_{pp}^{\Psi} +N_{pp}^{B\rightarrow \Psi})\cdot N_{coll}$. 

In calculations before, transport model successfully explains the nuclear
modification factor $R_{AA}$ for charmonium~\cite{trans1, trans2} and bottomonium~\cite{trans3},
and also the excited state double ratio $R_{AA}^{\psi^\prime}/R_{AA}^{J/\psi}$~\cite{trans4}
with centrality, with (2+1)D ideal hydrodynamics and boost-invariant approximation in
longitudinal direction. 
In this work, I employ the (3+1)D ideal hydrodynamics~\cite{hirano} and show the different 
mechanisms of $J/\psi$ production at different rapidity regions.

\section{IV. rapidity dependence}
\begin{figure}[htb]
{\includegraphics[width=0.48\textwidth]{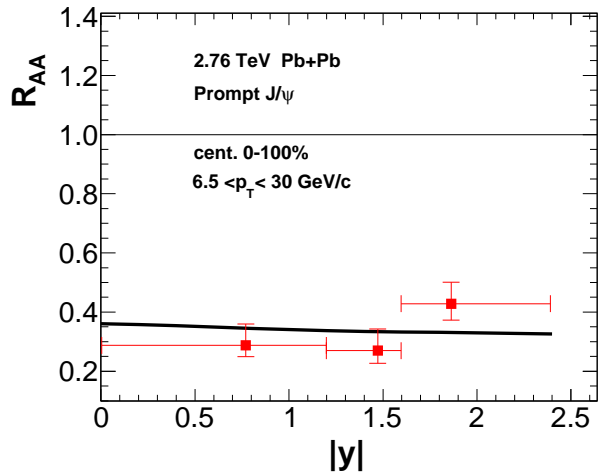}
\caption{(Color online) The prompt $J/\psi$ nuclear modification factor
as a function of rapidity for $\sqrt{s_{NN}}=2.76$ TeV Pb+Pb collisions.
The data is from the CMS Collaboration~\cite{Chatrchyan:2012np}.}
\label{pt6}}
\end{figure}

In large transverse momentum bin $6.5<p_T<30$ GeV/c, charmonium produced from parton 
hard scatterings at the nucleus colliding time ($\tau=0$) dominates the final total yields of $J/\psi$. 
The regenerated charmonium are produced from the recombination of 
thermalized charm quarks and their transverse momentum
is much smaller ($\langle p_T^\mathrm{regeneration}(J/\psi)\rangle \sim 2$ GeV/c). 
Therefore the contribution of charmonium regeneration 
is almost zero in Fig.\ref{pt6}~\cite{Zhao:2007hh}. 
The line includes both cold and
hot nuclear matter effects. Cold nuclear matter effects show weak rapidity dependence.
For hot nuclear matter effects, 
the evolutions of the medium are similar at different rapidity, 
due to the boost invariance of the hot medium in central rapidity region. This
results in a similar charmonium suppression at different rapidity and so the flat
feature of $R_{AA}^{J/\psi}(y)$.

\begin{figure}[htb]
{\includegraphics[width=0.48\textwidth]{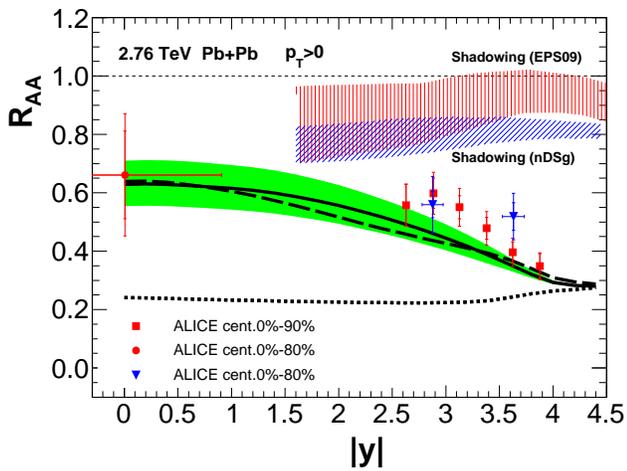}
\caption{(Color online) The inclusive $J/\psi$ nuclear modification factor
as a function of rapidity for $\sqrt{s_{NN}}=2.76$ TeV Pb+Pb collisions. 
Theoretical calculations are in centrality 0-90\%. 
Dotted line includes initially produced charmonium and non-prompt charmonium (B decay), 
with both cold and hot nuclear matter effects. Black solid line and 
black dashed line are the total yields of charmonium with both cold and hot nuclear matter effects. 
They correspond to the cross sections of solid line and dashed line respectively in Fig.\ref{figJpsiSigma}. 
For the black solid line, I also shift ${d\sigma^{pp}_{c{\bar c}}\over dy}(y)$ with 10\% to consider the 
the uncertainty of charm pair cross section, which results in the green color region of $R_{AA}$. 
Results of EPS09 and nDSg models are cited from~\cite{midpt1}. 
The experimental data are from the ALICE Collaboration~\cite{midpt1,pbdata}.}
\label{pt0}}
\end{figure}

The situation is so different when coming to the entire $p_T$ bin in Fig.\ref{pt0}. 
Cold nuclear matter effects can not explain the 
decreasing tendency of $R_{AA}^{J/\psi}$ in forward rapidity region $2.5<|y|<4$. 
Even shadowing effect shows strong rapidity dependence in p+Pb collisions, 
it shows very weak rapidity dependence in Pb+Pb collisions, 
see the results of EPS09 and nDSg models in Fig.\ref{pt0}. 
The rapidity dependence of $J/\psi$ observables in Pb+Pb collisions at 
LHC energies are mainly caused by hot medium effects. 
In $p_T>0$ bin in Fig.\ref{pt0}, 
charmonium regeneration 
contributes more than half of the final total yields, and 
the decrease 
of regenerated charmonium yields results in the decreasing tendency of $R_{AA}(y)$. 

Experimental data gives the inclusive $J/\psi$ yield which includes non-prompt $J/\psi$ from B hadron decay, and 
it contributes about 10\% to the final inclusive yield of $J/\psi$.
Instead of calculating each centralities of the experimental data in
Fig.\ref{pt0}, I do the theoretical calculations in centrality 0-90\%.
The total nuclear modification factor is 
$R_{AA}(y)=R_{AA}^{\mathrm{init}}(y)+R_{AA}^{\mathrm{rege}}(y)+R_{AA}^{B\to \Psi}(y)$, see Eq.(\ref{raaEq}). 
In central rapidity region, the produced hot medium is boost invariant.
This results in a flat tendency of $R_{AA}^{\mathrm{init}}(y)$ in Fig.\ref{pt0}, similar with the line in Fig.\ref{pt6}.
In forward rapidity region, with rapidity increasing,
the temperature of the medium decreases, and so $R_{AA}^{\mathrm{init}}(y)$ 
increases a little with rapidity (dotted line). 
The charm pair cross section decreases with rapidity (see Fig.\ref{figJpsiSigma}). 
It makes $R_{AA}^{\mathrm{rege}}(y)$ decreases with rapidity, and results in the decreasing tendency of $R_{AA}(y)$ at forward 
rapidity region. 
Here, I give two results of $R_{AA}(y)$ with different 
parametrization of ${d\sigma_{c\bar c}^{pp}\over dy}(y)$ and ${d\sigma_{J/\psi}^{pp}\over dy}(y)$. 
The form of ${d\sigma^{pp}_{c{\bar c}}\over dy}(y)=A\exp(-y^4/(2B^2))$ 
decrease slowly with rapidity in central rapidity region $|y|<2$, and decrease rapidly in forward rapidity region $2.5<|y|<4$. 
It results in the 
flat tendency of total nuclear modification factor $R_{AA}$ in central rapidity region, and the strong decreasing tendency in 
forward rapidity region. Black dashed line is for $R_{AA}$ with the charm pair 
cross section ${d\sigma^{pp}_{c{\bar c}}\over dy}(y)=A\exp(-y^2/(2B^2))$.

Considering the large error-bar of total charm pair cross section $\sigma_{c\bar c}^{pp}(\mathrm{total})$,
I shift ${d\sigma_{c\bar c}^{pp}\over dy}(y)$ up and down with $10\%$ to
consider the uncertainty, It results in the uncertainty of $R_{AA}(y)$ shown as green color region in Fig.\ref{pt0}.
\begin{figure}[htb]
{\includegraphics[width=0.48\textwidth]{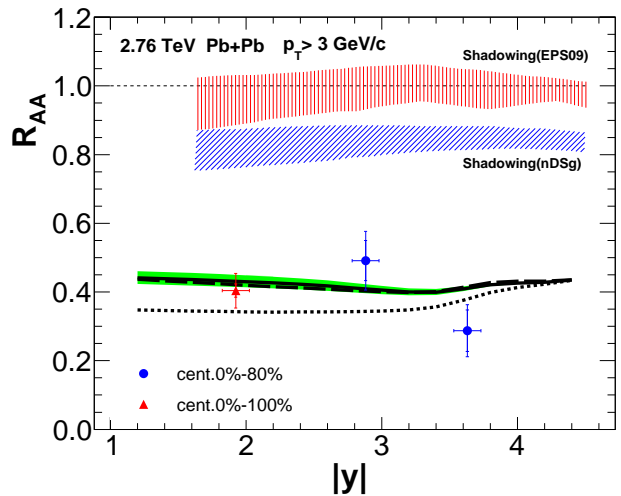}
\caption{(Color online) The inclusive $J/\psi$ nuclear modification factor in $p_T>3$ GeV/c region 
as a function of rapidity for $\sqrt{s_{NN}}=2.76$ Pb+Pb collisions.
Theoretical calculations are in centrality 0-90\%.
Black dotted line, dashed line, solid line 
and green color region are the same with those in Fig.\ref{pt0}, but in $p_T>3$ GeV/c. 
Results of EPS09 and nDSg models are cited from~\cite{midpt1}. 
The experimental data are from the ALICE Collaboration~\cite{midpt1}.
}
\label{pt3}}
\end{figure}

This decreasing tendency of $R_{AA}(y)$ is not so obvious in the data of the middle $p_T$ bin of $J/\psi$ in Fig.\ref{pt3},
with only three data points and relatively large error bar. 
In this $p_T$ bin, the 
fraction of B decay contribution in the final inclusive $J/\psi$ yield is about 15\%~\cite{Chatrchyan:2012np}.
In Fig.\ref{pt3}, there
is also a flat region of $R_{AA}$ in middle rapidity. A small decreasing tendency of $R_{AA}$ in $2<|y|<3.5$ is
caused by the decrease of charmonium regeneration. As regenerated charmonium is from the 
recombination of thermalized charm and anti-charm quarks, they are mainly distributed in low $p_T$ bin. 
Therefore in middle $p_T$ bin in Fig.\ref{pt3}, the decrease of regenerated charmonium yield only results 
in a weak decreasing tendency of total nuclear modification factor $R_{AA}$.  
In $y\ge 3.5$, 
temperature of the medium decreases
faster with rapidity, due to the end of boost invariance of hot medium. 
This gives less suppression of initially 
produced charmonium, and increase the $R_{AA}^{\mathrm{init}}(y)$ a little, see the dotted line. The competition between 
the increase of $R_{AA}^{\mathrm{init}}(y)$ and the decrease of $R_{AA}^{\mathrm{rege}}(y)$ results in the behavior of $R_{AA}(y)$ in 
$3<|y|<4$ region. Green color region is caused by the uncertainty of ${d\sigma_{c\bar c}^{pp}\over dy}(y)$. 
With only shadowing effects, $R_{AA}$ are far above the experimental data, please see Fig.\ref{pt3}. 

\begin{figure}[htb]
{\includegraphics[width=0.48\textwidth]{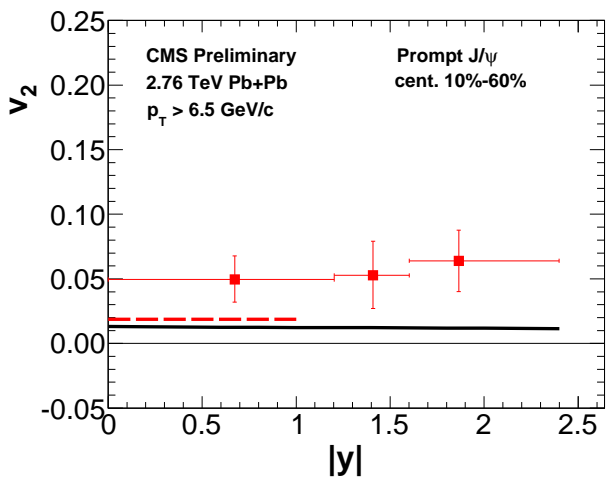}
\caption{(Color online) The elliptic flow of prompt $J/\psi$
as a function of rapidity for $\sqrt{s_{NN}}=2.76$ TeV Pb+Pb collisions.
The black line is for smooth (3+1)D ideal hydro, and red dashed line is for smooth (2+1)D viscous hydro.
The data is from the CMS Collaboration~\cite{v2CMS1}.}
\label{v2}}
\end{figure}

Even the rapidity dependence of prompt and inclusive $J/\psi$ nuclear modification factor
$R_{AA}$ are well described at different transverse momentum bins, the 
theoretical calculations underestimate the elliptic flow of $J/\psi$ 
in Fig.\ref{v2}. 
In $p_T>6.5$ GeV/c bin, initially produced $J/\psi$ dominates
the final yield. They are produced isotropically. However, the hot medium is anisotropic in centrality 0-60\%, and 
so the suppression of charmonium is different when they move in different 
directions in the hot medium. This results in small elliptic flow of $J/\psi$. 
In Fig.\ref{v2}, I employ both smooth (3+1)D ideal hydro (black solid line) 
and smooth (2+1)D viscous hydro (red dashed line) for the theoretical 
calculations. Both of them underestimate the experimental data. 
In the transport model, $J/\psi$ is assumed to move with constant velocity in QGP due to its large mass, this may 
underestimate its collective flows in high $p_T$ bin. New mechanism of interactions between $J/\psi$ and QGP may 
give additional enhancement of $v_2$ in Fig.\ref{v2}~\cite{Kopeliovich:2014una}.

\begin{figure}[htb]
{\includegraphics[width=0.48\textwidth]{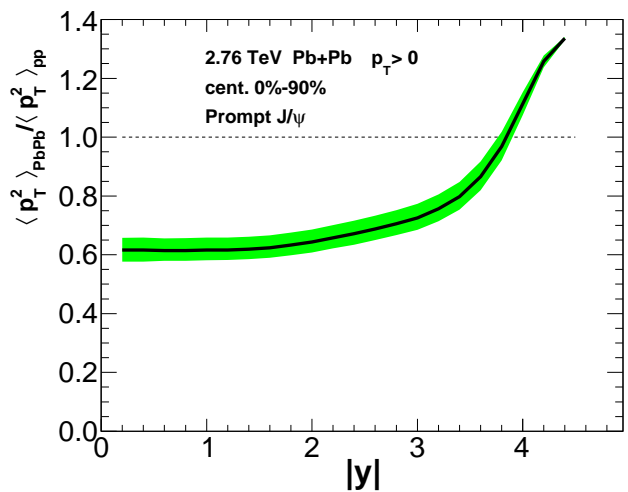}
\caption{(Color online) The mean transverse momentum square ratio $r_{AA}$ of
prompt $J/\psi$ as a function of rapidity for $\sqrt{s_{NN}}=2.76$ TeV Pb+Pb collisions.
Theoretical calculations are in the centrality 0-90\%. 
Black solid line and green color region are the same with those in Fig.\ref{pt0}. 
}
\label{pt2fig}}
\end{figure}

In order to distinguish the $J/\psi$ production mechanisms at different rapidity,
I also calculate the ratio of $J/\psi$ mean transverse momentum square in nucleus-nucleus
collisions to that in proton-proton collisions $r_{AA}=\langle p_T^2\rangle_{PbPb}/\langle p_T^2\rangle_{pp}$.
The mean transverse momentum of charmonium from parton hard 
scatterings (charmonium initial production) is much larger. The regenerated 
charmonium is from the recombination of thermalized charm and 
anti-charm quarks, and so the mean transverse momentum of charmonium is much smaller. 
In central rapidity region of Fig.\ref{pt2fig}, The yield of regenerated charmonium dominates 
the final total yields of charmonium, due to the large charm pair cross section ${d\sigma_{c\bar c}^{pp}\over dy}(y)$. 
It pulls down the mean transverse momentum square of final total $J/\psi$, 
and makes $r_{AA}<1$. With rapidity increasing, 
the fractions of charmonium yields from regeneration decrease. This 
increases the $\langle p_T^2\rangle_{PbPb}$. Also, the temperature of hot 
medium is smaller in forward rapidity region, which gives less suppression of charmonium. 
The fractions of charmonium initial production increases 
in final total yields. Unlike $R_{AA}$, both of above effects increase the $r_{AA}$, 
which makes $r_{AA}$ more sensitive to the mechanisms of charmonium production in hot medium. 
In $|y|>3.8$, $\langle p_T^2\rangle_{AA}$ is larger than 
$\langle p_T^2\rangle _{pp}$. This is caused by the leakage effect: 
charmonium with larger momentum suffer less suppression in hot medium, which 
increases the mean transverse momentum square of final total $J/\psi$ in Pb+Pb collisions and makes $r_{AA}>1$. 

\section{V. conclusion}
In summary, I employ both (3+1) dimensional Boltzmann-type transport equation and hydrodynamics
to calculate the detailed rapidity dependence of $J/\psi$ nuclear modification factor $R_{AA}(y)$
and the ratio of mean transverse momentum square $r_{AA}(y)$ at LHC $\sqrt{s_{NN}}=2.76$ TeV Pb+Pb collisions. 
Different from p+Pb collisions, 
in Pb+Pb collisions cold nuclear matter effects are almost independent of rapidity.
Hot medium suppression from gluon dissociation shows weak rapidity dependence,
due to boost invariance of the medium in central rapidity region.
Both the strong decreasing tendency of $R_{AA}$ and increasing tendency of $r_{AA}$ in forward
rapidity region are caused by the rapid decrease of
regeneration contribution. 
The ratio of mean transverse momentum square $r_{AA}$
seems more sensitive to the mechanisms of $J/\psi$ production.
Theoretical calculations explain the experimental data at high and middle $p_T$ bin and also the strong
decreasing tendency of $R_{AA}(y)$ in entire $p_T$ bin, but have some discrepancy with the data
of $J/\psi$ elliptic flow in Fig.\ref{v2}. 
More precise data and theoretical study are needed to understand the behavior of heavy flavor in the 
deconfined matter.

\appendix {\bf Acknowledgement}: Author thanks Dr.Fuming Liu in CCNU 
for helpful discussions in the beginning of this work, and Dr.Pengfei Zhuang and Dr.Kai Zhou 
for the good suggestions in this work.


\begin{thebibliography}{20}
\bibitem{matsui} T.Matsui and H.Satz, Phys. Lett. {\bf B178}, 416(1986).
\bibitem{Thews:2000rj} R.~L.~Thews, M.~Schroedter and J.~Rafelski, Phys.\ Rev.\ C {\bf 63}, 054905 (2001). 
\bibitem{Grandchamp:2002wp} L.~Grandchamp and R.~Rapp, Nucl.\ Phys.\ A {\bf 709}, 415 (2002). 
\bibitem{Andronic:2003zv} A.~Andronic, P.~Braun-Munzinger, K.~Redlich and J.~Stachel, Phys.\ Lett.\ B {\bf 571}, 36 (2003). 
\bibitem{Andronic:2011yq} A.~Andronic, P.~Braun-Munzinger, K.~Redlich and J.~Stachel, J.\ Phys.\ G {\bf 38}, 124081 (2011). 
\bibitem{RHIC1} A.Adare {\it et al.}, [PHENIX Collaboration], Phys. Rev. Lett. {\bf 98}, 232301(2007).
\bibitem{Liu:2009wza} Y.~Liu, Z.~Qu, N.~Xu and P.~Zhuang, J.\ Phys.\ G {\bf 37}, 075110 (2010).
\bibitem{Zhao:2008pp} X.~Zhao and R.~Rapp, Eur.\ Phys.\ J.\ C {\bf 62}, 109 (2009). 
\bibitem{Krouppa:2015yoa} B.~Krouppa, R.~Ryblewski and M.~Strickland, Phys.\ Rev.\ C {\bf 92}, no. 6, 061901 (2015).
\bibitem{Manceau:2013} L.Manceau, [ALICE Collaboration], EPJ Web Conf.\  {\bf 60}, 13002 (2013).
\bibitem{midpt1} B.Abelev {\it et al.}, [ALICE Collaboration], Phys. Rev. Lett. {\bf, 109}, 072301(2012).
\bibitem{Andronic:2007bi} A.~Andronic, P.~Braun-Munzinger, K.~Redlich and J.~Stachel, Phys.\ Lett.\ B {\bf 652}, 259 (2007). 
\bibitem{trans1} X.~l.~Zhu, P.~f.~Zhuang and N.~Xu, Phys.\ Lett.\ B {\bf 607}, 107 (2005). 

\bibitem{Liu:2009gx} Y.~Liu, N.~Xu and P.~Zhuang, Nucl.\ Phys.\ A {\bf 834}, 317C (2010). 
\bibitem{Grandchamp:2001pf} L.~Grandchamp and R.~Rapp, Phys.\ Lett.\ B {\bf 523}, 60 (2001). 
\bibitem{Grandchamp:2003uw} L.~Grandchamp, R.~Rapp and G.~E.~Brown, Phys.\ Rev.\ Lett.\  {\bf 92}, 212301 (2004). 
\bibitem{Zhao:2010nk} X.~Zhao and R.~Rapp, Phys.\ Rev.\ C {\bf 82}, 064905 (2010). 
\bibitem{OPE1} M.E.Peskin, Nucl. Phys. {\bf B156}, 365(1979).
\bibitem{OPE2} G.Bhanot, M.E.Peskin, Nucl. Phys. {\bf B156}, 391(1979).
\bibitem{radius2} H.Satz, J. Phys. {\bf G32}, R25(2006).
\bibitem{elastic1} B.~Chen, K.~Zhou and P.~Zhuang, Phys.\ Rev.\ C {\bf 86}, 034906 (2012).
\bibitem{Thews:2000fd} R.~L.~Thews, M.~Schroedter and J.~Rafelski, J.\ Phys.\ G {\bf 27}, 715 (2001). 
\bibitem{rege1} L.Yan, P.Zhuang, N.Xu, Phys. Rev. Lett {\bf 97}, 232301(2006).
\bibitem{Zhao:2011cv} X.~Zhao and R.~Rapp, Nucl.\ Phys.\ A {\bf 859}, 114 (2011). 
\bibitem{Du:2015wha} X.~Du and R.~Rapp, Nucl.\ Phys.\ A {\bf 943}, 147 (2015). 
\bibitem{trans2} K.~Zhou, N.~Xu, Z.~Xu and P.~Zhuang, Phys.\ Rev.\ C {\bf 89}, no. 5, 054911 (2014).
\bibitem{Zhao:2012gc} X.~Zhao, A.~Emerick and R.~Rapp, Nucl.\ Phys.\ A {\bf 904-905}, 611c (2013). 
\bibitem{ctherm1} A.Adare {\it et al.}, [PHENIX Collaboration], Phys. Rev. Lett. {\bf 98}, 172301(2007).
\bibitem{Greco:2003vf} V.~Greco, C.~M.~Ko and R.~Rapp, Phys.\ Lett.\ B {\bf 595}, 202 (2004). 
\bibitem{Eskola:1998df} K.~J.~Eskola, V.~J.~Kolhinen and C.~A.~Salgado, Eur.\ Phys.\ J.\ C {\bf 9}, 61 (1999). 
\bibitem{Vogt:2010aa} R.~Vogt, Phys.\ Rev.\ C {\bf 81}, 044903 (2010). 
\bibitem{Vogt:2004dh} R.~Vogt, Phys.\ Rev.\ C {\bf 71}, 054902 (2005). 
\bibitem{cronin1} S.Gavin, M.Gyulassy, Phys. Lett. {\bf B214}, 241(1988).
\bibitem{Hufner:2001tg} J.~Hufner and P.~f.~Zhuang, Phys.\ Lett.\ B {\bf 515} (2001) 115. 
\bibitem{Thews:2005vj} R.~L.~Thews and M.~L.~Mangano, Phys.\ Rev.\ C {\bf 73}, 014904 (2006). 
\bibitem{phenix1} A.Adare, {\it et al.}, [PHENIX Collaboration], Phys. Rev. Lett. {\bf 98}, 232002(2007).
\bibitem{Abelev:2012kr} B.~Abelev {\it et al.} [ALICE Collaboration], Phys.\ Lett.\ B {\bf 718}, 295 (2013). 
\bibitem{Chatrchyan:2012np} S.~Chatrchyan {\it et al.} [CMS Collaboration], JHEP {\bf 1205}, 063 (2012). 
\bibitem{Andronic:2006ky} A.~Andronic, P.~Braun-Munzinger, K.~Redlich and J.~Stachel, Nucl.\ Phys.\ A {\bf 789}, 334 (2007).
\bibitem{cpair1} [ALICE Collaboration], JHEP {\bf 1207}, 191(2012).
\bibitem{Cacciari:2001td} M.~Cacciari, M.~Greco and P.~Nason, JHEP {\bf 9805}, 007 (1998); M.~Cacciari, S.~Frixione and P.~Nason, JHEP {\bf 0103}, 006 (2001). 
\bibitem{trans3} Y.~Liu, B.~Chen, N.~Xu and P.~Zhuang, Phys.\ Lett.\ B {\bf 697}, 32 (2011).
\bibitem{trans4} B.~Chen, Y.~Liu, K.~Zhou and P.~Zhuang, Phys.\ Lett.\ B {\bf 726}, 725 (2013).
\bibitem{hirano} T.Hirano, Phys. Rev. {\bf C65}, 011901(2001).
\bibitem{Zhao:2007hh} X.~Zhao and R.~Rapp, Phys.\ Lett.\ B {\bf 664}, 253 (2008). 
\bibitem{pbdata} [ALICE Collaboration], Nucl. Phys. {\bf A904}, 202c(2013)
\bibitem{Kopeliovich:2014una} B.~Z.~Kopeliovich, I.~K.~Potashnikova, I.~Schmidt and M.~Siddikov, Phys.\ Rev.\ C {\bf 91}, no. 2, 024911 (2015). 
\bibitem{v2CMS1} [CMS Collaboration], CMS-PAS-HIN-12-001(2013).

\end{thebibliography}
\end{document}